\documentclass[a4paper]{article}

\usepackage{17lomcon}        % Proceedings volume layout metrics
\usepackage{cite}             % Smart range citations
\usepackage{epsfig}           % Encapsulated PostScript figure inclusion
\usepackage{amsmath,amssymb}
\usepackage{color}
\usepackage{comment}
\usepackage{float}
\usepackage{cite}

\begin{document}

%%%%%%%%%%%%%%%%%%%%%%%%%%%%%%%%%%%%%%%%%%%%%%%%%%%%%%%%%%%%%%%%%%
% The preamble of the paper
%%%%%%%%%%%%%%%%%%%%%%%%%%%%%%%%%%%%%%%%%%%%%%%%%%%%%%%%%%%%%%%%%%

\title{LHC OPTICS AND ELASTIC SCATTERING  MEASURED BY THE TOTEM EXPERIMENT}

\author{F. Nemes, for the TOTEM Collaboration$^{1,2}$ \email{fnemes@cern.ch}\vspace{2mm}}

\affiliation{$^{1}$CERN, Geneva, Switzerland \\$^{2}$Wigner RCP, H-1121 Budapest XII, Konkoly-Thege 29-33, Hungary \vspace{2mm}}

% You may repeat \author and \affiliation as many times as necessary!

\date{}
% Print it out!
\maketitle

%%%%%%%%%%%%%%%%%%%%%%%%%%%%%%%%%%%%%%%%%%%%%%%%%%%%%%%%%%%%%%%%%%
% The preamble of the paper
%%%%%%%%%%%%%%%%%%%%%%%%%%%%%%%%%%%%%%%%%%%%%%%%%%%%%%%%%%%%%%%%%%

\begin{abstract}
	The TOTEM experiment at the LHC has measured proton-proton elastic scattering in dedicated
	runs at $\sqrt{s}=7$ and 8 TeV centre-of-mass LHC energies. The proton-proton total cross-section $\sigma_{\rm tot}$ has
	been derived for both energies using a luminosity independent method. TOTEM has excluded a purely exponential differential
	cross-section for elastic proton-proton scattering with significance greater than 7$\sigma$ in the $|t|$ range from 0.027 to 0.2 GeV$^{2}$
	at $\sqrt{s}=8$~TeV.
\end{abstract}

\section{Introduction}

\begin{figure}[H]
		\centering
    		\includegraphics[trim = 145mm 97mm 88mm 52mm, clip, width=0.8\textwidth]{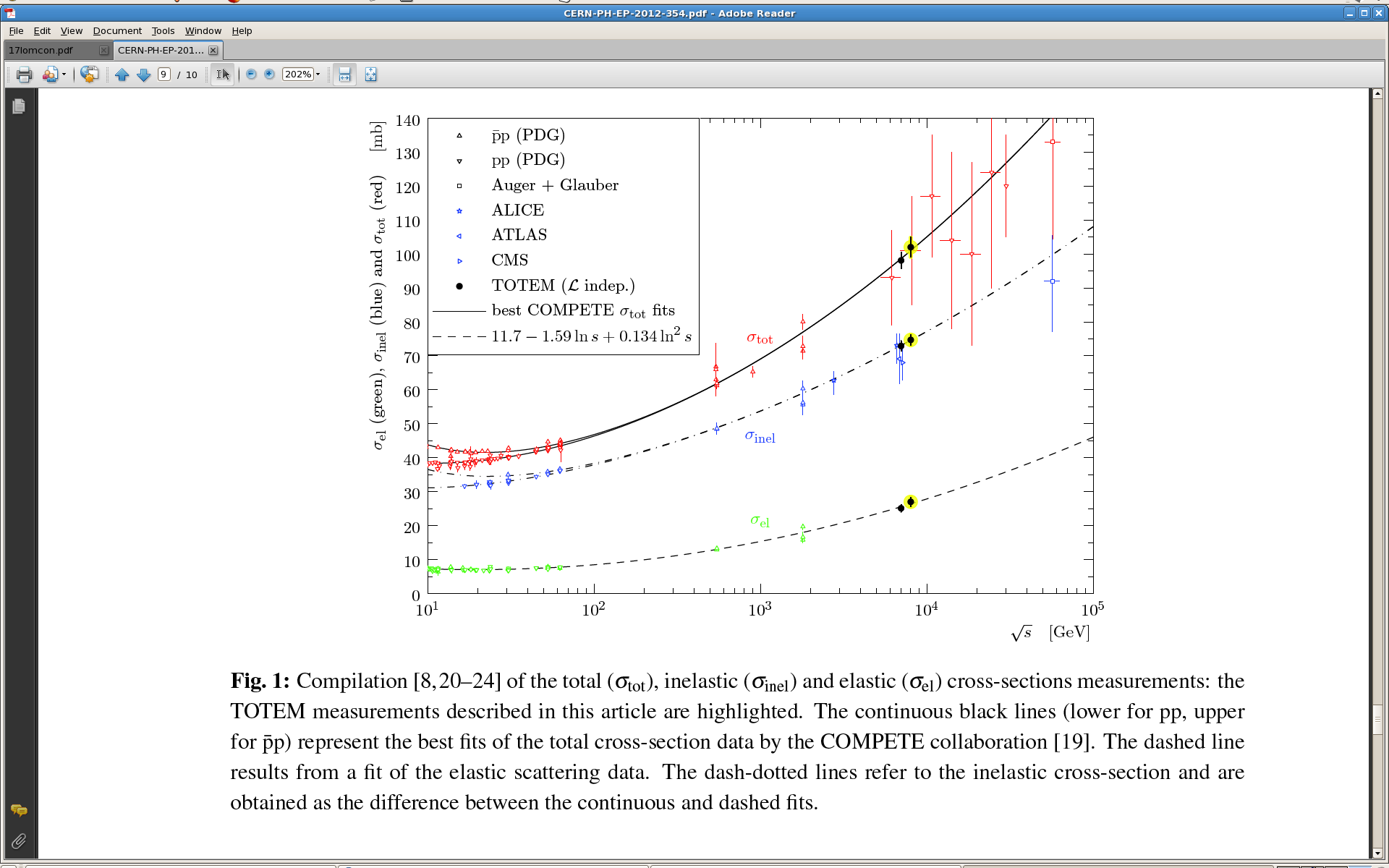}\vspace{-3mm}
		\caption{A compilation of total, inelastic and elastic pp cross-section measurements. The black points indicate the TOTEM measurements at $\sqrt{s}=7$ and 8~TeV using the luminosity independent
		method~\cite{Antchev:2013iaa,Antchev:2013paa, Agashe:2014kda, Abelev:2012sea, Chatrchyan:2012nj, ATLASinelastic}.}
		\label{elasticscatteringsummary}
	\end{figure}

	The TOTEM (TOTal cross section, Elastic scattering and diffraction dissociation Measurement at the LHC) experiment has been designed to measure the total proton-proton (pp) cross-section, elastic scattering and
	diffractive processes at the LHC~\cite{Anelli:2008zza}, see Fig.~\ref{elasticscatteringsummary}.

	The experimental apparatus of TOTEM is composed of three subdetectors: the Roman Pots (RP), the T1 and T2 inelastic forward telescopes. The detectors are placed symmetrically on both sides of the Interaction
	Point 5 (IP5), which is shared with the CMS experiment.

	The RPs are moveable beam-pipe insertions, hosting edgeless silicon detectors to detect leading protons scattered at very small angles.
 	In order to maximize the acceptance of the experiment for elastically scattered protons, the RPs are able to approach the beam center to a transverse
	distance as small as 1 mm. The alignment of the RPs is optimized by reconstructing common tracks going through the overlap between the vertical and
	horizontal RPs~\cite{Anelli:2008zza,Antchev:2011zz}.

	Before the LHC long shutdown one (LS1) the RPs, used for measurements, were located at distances of 215--220 m from IP5~\cite{Anelli:2008zza}. The actual layout, i.e., after the LHC LS1, is different in RP
	location and quantity. The RP stations previously installed at $\pm$147~m, from IP5, have been relocated to $\pm$210~m. Moreover, two newly designed horizontal RPs have been installed
	between the two units of the $\pm220$~m station~\cite{TOTEM:2013iga,Albrow:2015ois}.

\section{Elastic scattering and total cross-section $\sigma_{\rm tot}$ measurements}

\begin{figure}[H]
		\centering
    		\includegraphics[width=0.49\textwidth]{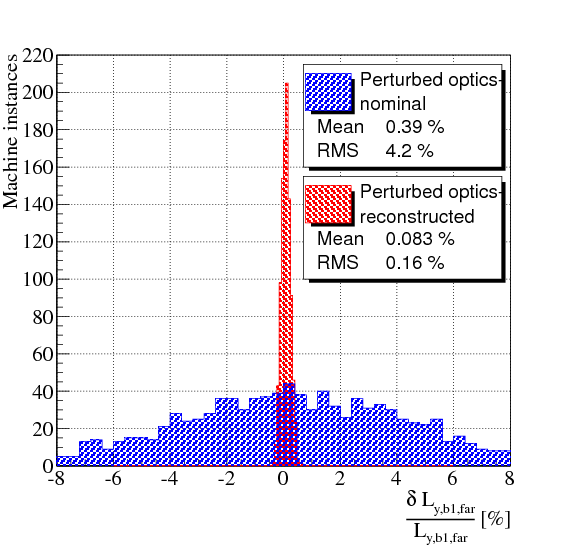}
    		\includegraphics[width=0.49\textwidth]{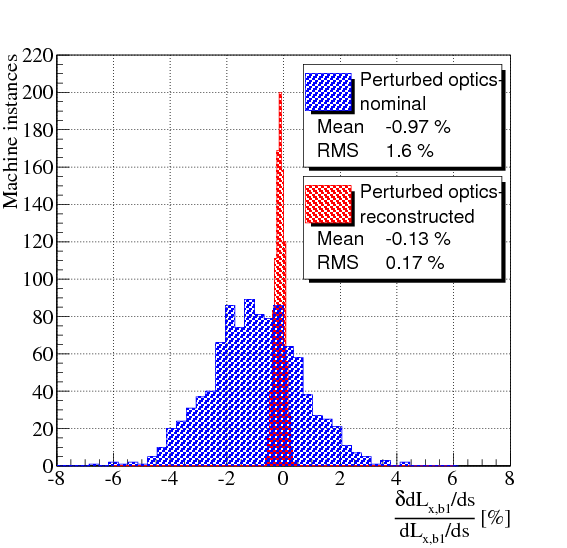}
		\caption{The MC error distribution of $\beta^{*}=3.5$ m optical functions $L_{y}$ and ${\rm d} L_{x}/{\rm d} s$ for Beam 1 at $\sqrt{s}=7$~TeV, before and after optics estimation.}
		\label{optical_functions}
	\end{figure}

        For each tagged elastic event the four-momentum transfer squared~$t$ is reconstructed using the LHC optical functions, characterized with the so-called
	betatron amplitude at IP5~$\beta^{*}$~\cite{Anelli:2008zza}. The TOTEM experiment developed a novel experimental method to estimate the optical functions at the RP locations,
	using the measured elastically scattered protons~\cite{Antchev:2014voa,Antchev:2011vs,Antchev:2013gaa,Antchev:2013haa}, see Fig.~\ref{optical_functions}.\vspace{-4mm}

	The total inelastic rate $N_{\rm inel}$, measured by the T1 and T2 telescopes, and the total nuclear elastic rate $N_{\rm el}$
	with its extrapolation to zero four-momentum transfer squared~$t=0$ are combined with the optical theorem to obtain the total cross-section
	in a luminosity, $\mathcal{L}$, independent way
	\begin{equation}
		\sigma_{\rm tot}=\frac{16\pi}{1+\rho^{2}}\cdot\left.\frac{{\rm d}N_{\rm el}}{{\rm d}t}\right|_{t=0}\cdot(N_{\rm el}+N_{\rm inel})^{-1}.
	\end{equation}
	The measured ratio of the elastic and inelastic rates $N_{\rm el} / N_{\rm inel}$ allows for the determination of the elastic and inelastic cross-sections as well~\cite{Anelli:2008zza,Antchev:2013haa}.

	The TOTEM experiment determined the total pp cross-section at $\sqrt{s}=7$~TeV using the luminosity independent method~\cite{Antchev:2013iaa}, which was shown to be consistent with the total cross-sections
	measured in independent ways, see Table~\ref{totalcrosssections}. The elastic and inelastic cross-sections were found to be $\sigma_{\rm el}=25.1 \pm 1.1$~mb and $\sigma_{\rm inel}=72.9 \pm 1.5$~mb.

	The measurement was repeated at $\sqrt{s}=8$~TeV, yielding to $\sigma_{\rm tot}=101.7\pm2.9$~mb, $\sigma_{\rm el}=27.1\pm1.4$~mb and $\sigma_{\rm inel}=74.7\pm1.7$~mb~\cite{Antchev:2013paa}. A compilation of the results is shown in Fig.~\ref{elasticscatteringsummary},
	which also demonstrates that the observed cross-sections are in agreement with low-energy data and cosmic ray results as well~\cite{Abelev:2012sea, Chatrchyan:2012nj, ATLASinelastic,Agashe:2014kda}.

	\begin{table}[H]
		\centering
		\begin{tabular}{|c|c|c|c|c|} \hline
			Method				& $\mathcal{L}$ independent~\cite{Antchev:2013iaa}	&~\cite{Antchev:2011vs}		&~\cite{Antchev:2013gaa}	&~\cite{Antchev:2013gaa}\\ \hline
			$\sigma_{\rm tot}$ [mb] 	& 98.0 $\pm$ 2.5					& 98.3 $\pm$ 2.8		& 98.6 $\pm$ 2.2		& 99.1 $\pm$ 4.3	\\ \hline
		\end{tabular}
		\caption{The total cross-section $\sigma_{\rm tot}$ results measured by the TOTEM experiment at $\sqrt{s}=7$~TeV with four different methods.}
		\label{totalcrosssections}
	\end{table}

\begin{figure}[h]
		\centering
    		\includegraphics[width=0.9\textwidth]{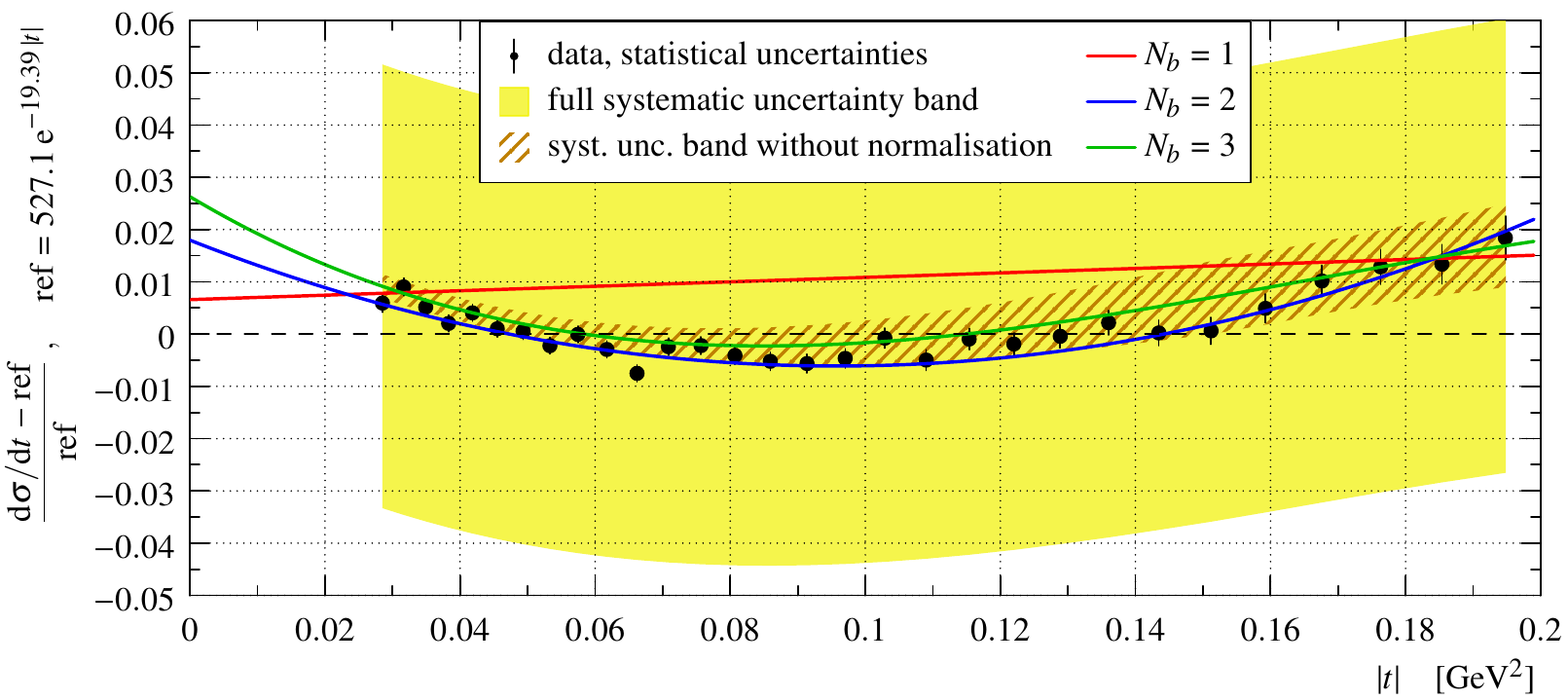}
		\caption{Differential cross-section measured at $\sqrt{s}=8$~TeV LHC energy plotted as relative difference from a reference exponential.
			The black dots represent data points with statistical uncertainty bars~\cite{Antchev:2015zza}.}
		\label{coulomb}
	\end{figure}

Thanks to a very-high statistics $\beta^{*}=90$~m data set at $\sqrt{s}=8$~TeV energy, the TOTEM experiment has excluded a purely exponential elastic pp differential cross-section~\cite{Antchev:2015zza}. The
exclusion's significance is greater than 7$\sigma$ in the $|t|$ range from 0.027 to 0.2 GeV$^{2}$, see Fig.~\ref{coulomb}. Using refined parametrizations for the extrapolation
to the optical point, $t=0$, yields total cross-section values $\sigma_{\rm tot}=101.5\pm2.1$~mb and $\sigma_{\rm tot}=101.9\pm2.1$~mb, compatible with the previous
TOTEM measurement.
	The TOTEM experiment performed its first measurement of elastic scattering in the Coulomb-nuclear interference (CNI) region~\cite{Coulombpaper}.
	The data have been collected at $\sqrt{s}=8$~TeV with a special beam optics of $\beta^{*}=1000$~m.
	The $\rho$ parameter was for the first time at LHC extracted via the Coulomb-nuclear interference, and
	was found to be~$\rho = 0.12\pm0.03$.\vspace{-2mm}
	
	\section{Conclusions}
	The TOTEM experiment has measured elastic pp scattering at $\sqrt{s}=$7 and 8~TeV. The 
	total, elastic and inelastic cross-sections have been derived for both energies using
	a luminosity independent method. TOTEM has excluded a purely
	exponential differential cross-section for elastic pp scattering with significance greater
	than 7$\sigma$ in the $|t|$ range from 0.027 to 0.2 GeV$^{2}$ at $\sqrt{s}=8$~TeV.
	Using $\beta^{*}=1000$~m optics at $\sqrt{s}=8$~TeV energy the $\rho$ parameter was for the first time at LHC extracted via the Coulomb-nuclear interference.

\section{Acknowledgement}\vspace{-1mm}

This work was supported by the institutions involved in the TOTEM collaboration and partially also by NSF (US), the Magnus Ehrnrooth Foundation (Finland), the 
Waldemar von Frenckell Foundation (Finland), the Academy of Finland, the Finnish Academy of Science and Letters 
(The Vilho, Yrjo and Kalle Vaisa la Fund) as well as by the Hungarian OTKA grant NK 101438.

\end{document}